\documentclass[aps,preprint,onecolumn,groupedaddress,nofootinbib,superscriptaddress]{revtex4-2}
\usepackage{graphicx}  % needed for figures
\usepackage{dcolumn}   % needed for some tables
\usepackage{bm}        % for math
\usepackage{amsmath,amssymb,amsxtra,bm,epsfig}   % for math
\usepackage{float}
\usepackage[dvipsnames]{xcolor}
\usepackage{xspace}
\usepackage{pifont}
\usepackage{soul}
\usepackage[]{changes}

 \def \vosc {{\rm{V_{osc}}}}

\usepackage{subfigure}
\usepackage{enumitem}

\setlist[enumerate]{nosep}

\usepackage[bookmarks, breaklinks, colorlinks,urlcolor=RoyalBlue, citecolor=RubineRed, 
linkcolor=ForestGreen]{hyperref}

\usepackage{mathtools}

\begin{document}

\title{%Life on the far side of $U_{\texttt{PMNS}}$
CP-conserving $SO(3)$ parameterization \\of the neutrino mixing matrix
}

\author{Jaros\l{}aw Duda}
\email{jaroslaw.duda@uj.edu.pl}
\affiliation{Jagiellonian University, Cracow, Poland}
 
\author{Janusz Gluza}
\email{janusz.gluza@us.edu.pl}
\affiliation{Institute of Physics, University of Silesia,  Katowice, Poland}

\author{Biswajit Karmakar}
\email{biswajit.karmakar@us.edu.pl}
\affiliation{Institute of Physics, University of Silesia,  Katowice, Poland}

\begin{abstract}
The pattern of neutrino mixing, usually parameterized by the Pontecorvo–Maki–Nakagawa– Sakata $U_{\rm PMNS}$ matrix, still remains a striking puzzle in particle physics.  $U_{\rm PMNS}$ is one of six possible products of multiplying three Euler matrices. Here we discuss the neutrino mixing matrix parameterization for three known flavours of neutrinos based on the SO(3) group represented by one three-dimensional rotation matrix $U_{\rm SO3}$  and the CP-conserving phases $\delta_{\rm CP}=0^{\circ}$ and $180^\circ$. The mixing matrix $U_{\rm SO3}$ with cyclic order of the  Lie group generators implies $\delta_{\rm{CP}}=180^\circ$ for clockwise rotation in three dimensions, a viable scenario for normal mass ordering.   We determine a range of rotation angles for $U_{\rm SO3}$ which deviate substantially from the almost maximal mixings in the standard  $U_{\rm PMNS}$ scenario, yielding  `democratic' values for the mixing angles. On the other hand, $\delta_{\rm CP}=0^{\circ}$ CP-conserving case supports near-maximal mixing scenario. With the fixed 
$\delta_{\rm CP}$ value, the $U_{\rm SO3}$ parameterization can be validated or falsified by the next generation of neutrinoless double beta decay experiments and puts a stringent constraint on the absolute neutrino mass.  The proposed  $U_{{\rm SO3}}$  neutrino mixing parameterization is suited for independent  CP-conserving neutrino oscillation experimental analysis.  
\end{abstract}
	%\pacs{}

    \maketitle

  \newpage  
\section{Introduction}

The simplest theory of massive neutrinos is based on the three-neutrino paradigm with the assumption that flavor states ($\nu_e,\nu_\mu,\nu_\tau$) are mixed with massive states ($\nu_1,\nu_2,\nu_3$) with definite masses ($m_1,m_2,m_3$), at least two of which are non-zero.  Several experiments around the globe on solar, atmospheric, reactor, and accelerator neutrino oscillations have provided compelling evidence for non-zero neutrino masses and non-trivial mixing. In the standard theory of neutrino oscillation, the three-neutrino mixing in the weak charged-current interaction can be written as
\begin{equation}
\nu_{\alpha} =\sum_{i=1}^3
U  _{\alpha i} 
{\nu_i}\;.
\label{class}
\end{equation}
where  $\alpha=e, \mu, \tau$ ($i=1,2,3$) stand for indices of flavour (mass) basis.  The standard parameterization (in the basis where charged lepton mass matrix is diagonal) of $U  _{\alpha i}$, the so-called Pontecorvo–Maki–Nakagawa–Sakata (\rm PMNS) unitary mixing matrix reads~\cite{Pontecorvo:1957qd,Maki:1962mu, Kobayashi:1973fv} 
	\begin{align}
  U_{\rm PMNS} &=R_x(\theta_{23})R_y(\theta_{13}, \delta_{\rm CP})R_z(\theta_{12})U_M(\alpha_1,\alpha_2)=U U_M(\alpha_1,\alpha_2)\label{upmns0}  \\
  &=\begin{pmatrix}
    1 & 0 & 0 \\
    0 & c_{23}  & {s_{23}} \\
    0 & -s_{23} & {c_{23}}
  \end{pmatrix}
  \begin{pmatrix}
 c_{13} & 0 & s_{13} e^{-i\delta_\text{\rm CP}} \\
    0 & 1 & 0 \\
    -s_{13} e^{i\delta_\text{\rm CP}} & 0 & c_{13}
  \end{pmatrix}
  \begin{pmatrix}
    c_{12} & s_{12} & 0 \\
    -s_{12} & c_{12} & 0 \\
    0 & 0 & 1
  \end{pmatrix} U_M \label{deltaCP} \\
  &=\begin{pmatrix}
c_{12} c_{13} & s_{12} c_{13} & s_{13} e^{-i\delta_{\text{CP}}} \\
-s_{12} c_{23} - c_{12} s_{23} s_{13} e^{i\delta_{\text{CP}}} & 
c_{12} c_{23} - s_{12} s_{23} s_{13} e^{i\delta_{\text{CP}}} & 
s_{23} c_{13} \\
s_{12} s_{23} - c_{12} c_{23} s_{13} e^{i\delta_{\text{CP}}} & 
-c_{12} s_{23} - s_{12} c_{23} s_{13} e^{i\delta_{\text{CP}}} & 
c_{23} c_{13}
\end{pmatrix}U_M
\label{upmns1}
\end{align}
Here $U_M$=diag$(e^{i\alpha_1},e^{i\alpha_2},1)$ stands for Majorana phase matrix, $c_{ij} \equiv \cos\theta_{ij}$, $s_{ij} \equiv \sin\theta_{ij}$, and the Euler rotation angles $\theta_{ij}$ can be taken without loss of generality from the first quadrant, $\theta_{ij} \in [0, \pi/2]$, and the Dirac CP phase $\delta_{\text{\rm CP}}$ and Majorana phases $\alpha_1,\alpha_2$ are in the range $ [0, 2\pi]$ \cite{10.1093/ptep/ptac097c14}. This choice of parameter regions is independent of matter effects~\cite{Gluza:2001de}, and the Majorana phases do not influence neutrino oscillation probabilities  \cite{Bilenky:1980cx, Giunti:2010ec}. In an equivalent   parameterization~\cite{Schechter:1980gr, Schechter:1980gk, Rodejohann:2011vc}, the lepton mixing matrix can be written in a `symmetrical' form where all three CP-violating phases are `physical'. 

In contrast with the standard form of $U_{\rm PMNS}$ given in Eq. (\ref{upmns1}),  here we propose a different view on neutrino mixing, read: rotations, which are based on the SO(3) symmetry group. Mathematically, the SO(3) group is a special orthogonal group in 3 dimensions and represents spatial rotations in three dimensions. It is a Lie group with a smooth manifold structure and continuous transformations. Every rotation can be uniquely represented by an orthogonal matrix with a determinant of 1, and these matrices form a group under matrix multiplication. 

The non-Abelian SO(3) group has already been explored in the neutrino theory in the context of discrete flavour symmetries and the construction of mass matrices from which mixings follow. For a review, see \cite{King:2013eh, Chauhan:2023faf}.  For explicit models incorporating SO(3) symmetry (with varied neutrino mass generation mechanisms) resulting in different mixing patterns, see \cite{Barbieri:1999km, King:2005bj, Antusch:2004xd, Wu:2007tq, Reig:2018ocz} and for analyses of spontaneous SO(3) breaking into discrete subgroups in the context of neutrino masses and mixing, see ~\cite{Etesi:1996urw, deMedeirosVarzielas:2005qg, Berger:2009tt, King:2018fke, Rachlin:2017rvm}.  The anarchy approach, on the other hand, assumes a structureless, randomly distributed neutrino mass matrix and uses the SO(3) Haar measure to predict near-maximal mixing in a basis-independent manner~\cite{Haba:2000be}. In turn,  a geometrical structure of the physical neutrino mixing matrix has been investigated in \cite{Flieger:2022ekj}, and the SO(3) surface area for the CP-conserving mixing case was determined. Furthermore, apart from the widely used PMNS picture given in Eq.~(\ref{upmns1}), there also exist different parameterizations of the neutrino mixing matrix, for a non-exhaustive list see ~\cite{Schechter:1980gr,Fritzsch:2001ty,Dattoli:2008zz,Gerard:2012ft,Zhukovsky:2016bee,Zhukovsky:2019eoy,Li:2005ir,Rodejohann:2003sc,Everett:2005ku,Rodejohann:2011vc,Merfeld:2014cha,Emmanuel-Costa:2015tca,Potter:2015pia,Zhukovsky:2016mon,Boriero:2017tkh,Davydova:2019aat,Zhukovsky:2019jzz} and references there-in.

Here, we directly apply the exponential representation of the SO(3) group to the description of the three-neutrino flavor mixing, independent of mass-generation
mechanism and underlying models, interpreting neutrino mixing as a compact spatial-like rotation governed by the Lie group generators, which results in a unique definition of mixing parameters.  We show that a single rotation, described by the mixing matrix  $U_{\rm SO3}$ and rooted in the SO(3) group theory, may account for the observed (CP-conserving) neutrino oscillation data.  Consequently, it also helps us to obtain a constrained prediction on the absolute neutrino mass being probed by tritium beta and neutrinoless double beta decay experiments with interesting phenomenological consequences.

In the next section, we consider order-dependent multiplication of Euler matrices, which brings the standard neutrino mixing $U_{\rm PMNS}$ parameterization (one of the six possible multiplication options), and the proposed $U_{\rm SO3}$ mixing parameterization based on the SO(3) group. In the section which follows we discuss the consequences of the proposed $U_{\rm SO3}$ parameterization and its possible application. We finish with conclusions and an outlook.

\section{Order dependence of neutrino mixing }
\subsection{Standard order dependent parameterization \label{standard}}

In the case of the standard  $U_{\rm PMNS}$ matrix parameterization given in Eq. (\ref{upmns1}), two of the three mixing angles, namely solar ($\theta_{12}$) and atmospheric ($\theta_{23}$), are found to be large, while the reactor ($\theta_{13}$)  mixing angle is relatively small. Neutrino oscillation data also constrain the two mass squared differences (solar and atmospheric) defined as $\Delta m_{21}^2 = m_2^2-m_1^2$ and $|\Delta m_{31}^2| = | m_3^2-m_1^2|$ respectively,  where $m_1, m_2, m_3$ are the masses of the three light neutrinos. The present NuFIT-6.0 global analysis\footnote{Other global fits are given by de Salas et al. ~\cite{deSalas:2020pgw,10.5281/zenodo.4726908} 
and Capozzi et al. ~\cite{Capozzi:2021fjo}. The small deviations among them are not important for the purpose of the present work.} gives ~\cite{esteban2025nufit}  
\begin{eqnarray}
&&  \Delta m^2_{21}=(7.30 - 7.68)\times10^{-5}\hspace{.1cm} \rm{eV}^2, \hspace{.5cm} |\Delta m^2_{31}|=(2.494 - 2.534)\times10^{-3}\hspace{.1cm} \rm{eV}^2, \nonumber  \\
 && \sin^2\theta_{12}=0.297-0.320, \hspace{.2cm} \sin^2\theta_{23}=0.457-0.487, \hspace{.2cm} \sin^2\theta_{13}=0.02494-0.02534,  \label{eq:oscdata}
\end{eqnarray}
for normal ordering (NO) of light neutrino mass and similar constraints for inverted ordering (IO).  Altogether, unravelling the leptonic flavour structure, in which neutrino mixing substantially differs from the quark sector, remains an outstanding problem in particle physics.  Currently, the Dirac CP violating phase $\delta_{\textrm{CP}}$  is estimated as $\delta_{\textrm{CP}}  = {212^\circ}^{+26^\circ}_{-41^\circ}$ at 1$\sigma$ \cite{esteban2025nufit}. However, there is still a tension and 2$\sigma$ disagreement between T2K and NO$\nu$A oscillation experiments in the measurement of $\delta_{\textrm{CP}}$~\cite{Chatterjee:2024kbn}. T2K prefers a value of  $\delta_{\rm{CP}}\simeq 1.5 \pi$  and NO$\nu$A indicating $\delta_{\rm{CP}} \simeq 0.9 \pi$ in the PMNS framework with normal mass ordering \cite{T2K:2019bcf, NOvA:2019cyt}.  Hence, the CP-conserving case with $\delta_{\rm {CP}}=180^\circ$ remains viable, particularly as NO$\nu$A's best-fit lies in close proximity. The recent combined T2K and NO$\nu$A collaborations analysis \cite{T2K:2025wet} are still not conclusive, leaving also a possibility for  $\delta_{\rm{CP}}=0$.  On the other hand,  $\alpha_1,\alpha_2$ phase factors are meaningful only if neutrinos are Majorana particles, an issue which still needs to be settled.  In case of Majorana neutrinos, to obtain $U\in \textrm{SO(3)}$ {\it we set these phases to zero}. By choosing the constraint $\alpha_{1,2} = 0$ we decided to take the most conservative path to remain within the SO(3) group. CP conservation for Majorana phases allows discrete values (0 or $\pi$, modulo 2$\pi$). To be more specific, following \cite{Bilenky:1987ty}, the nonzero eigenvalues of a real symmetric matrix can be either positive or negative $m'_{k}=\rho_k m_k$ where $m_k=|m'_k|$ and $\rho_k=\pm 1$, and the CP parity of the Majorana fields can be written as $\eta_{CP}(\chi_k)=i\rho_k$. 
With $\rho_k =+ 1$ and $-1$, CP-parities of Majorana neutrinos are opposite and some elements of the mixing matrix become pure complex, which has phenomenological consequences (destructive interferences are possible) which affect in particular neutrinoless double beta decay predictions and potential lepton flavour violation collider signals \cite{Gluza:2016qqv,Gluza:2015goa}.  
Detailed analyses of CP-parities for Majorana neutrinos, in particular with opposite CP-parities, go beyond this work. Such an analysis can be performed in analogy to collider studies for heavy Majorana neutrinos, with various signatures of CP-parities as given in \cite{Gluza:1995js,Gluza:1995ky,Gluza:1997ts}. 

Furthermore, there also exists an ambiguity on the solar mixing angle $\theta_{12}$ (involved in $R_z$ in Eq. (\ref{upmns0}))  arising from a degeneracy where potential non-standard interactions allow solar neutrino data to be explained by either the standard large mixing angle (LMA) solution (with $\sin^2\theta_{12}\simeq 0.31$) or the alternative Dark-LMA solution (with $\sin^2\theta_{12}\simeq 0.70$)~\cite{Miranda:2004nb,Farzan:2017xzy}.  Recently, a novel approach for examining the neutrino mixing matrix in Eq. (\ref{class}) has been proposed, 
where the unitarity of the lepton mixing matrix is tested without assuming a specific parameterization. The elements of the lepton mixing matrix can be extracted directly, without assuming a specific parametrization, from the energy dependence of the oscillation probabilities~\cite{Kitano:2025wpc}.

\begin{figure*}[h]
    \centering
        \includegraphics[width=.7\textwidth]{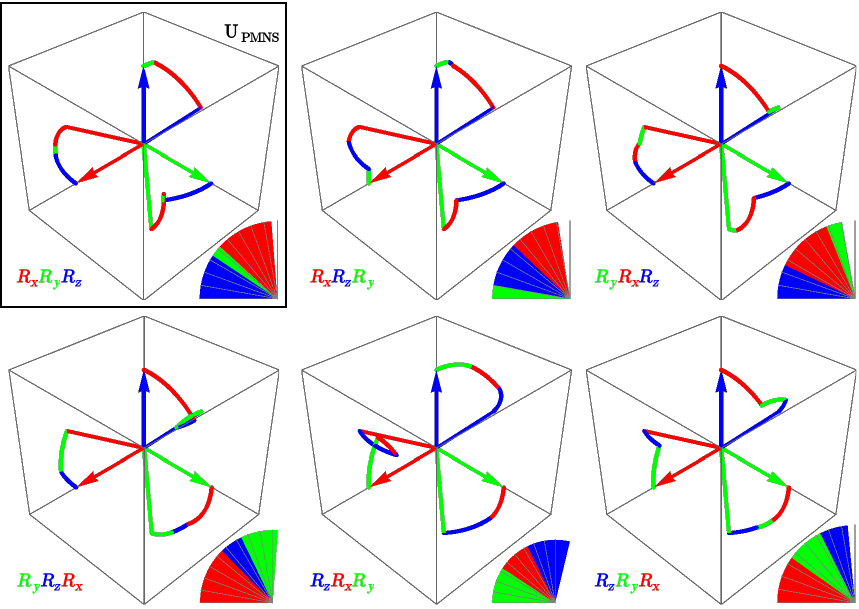}
        \caption{The $U$ matrix constructed from Euler angles written in (\ref{eul}) for 6 orderings. The framed subplot represents the standard \texttt{PMNS}  matrix order $R_x R_y R_z$.  
         The red-green-blue arc plots with $10^\circ$ marks visualise angles of performed rotations.  
        }
       \label{Euler}
\end{figure*}

In this work, we first scrutinize the parameterization of the $U$ mxing matrix given in Eq. (\ref{class}). The standard parameterization of the $U$ matrix in Eq.~(\ref{upmns1})  is one of six possible products of three-dimensional rotations for 3 flavors.
This situation is summarized in Fig.~\ref{Euler} for $\delta_{\textrm{CP}}=180^\circ$, a value which will be discussed in the next section. 
In Fig.~\ref{Euler}, the mixing angles are fixed with the best-fit values~\cite{esteban2025nufit} connected with standard $R_xR_yR_z$ order of rotations to be: $\theta_{23}=43.3^\circ, \theta_{13}=8.56^\circ,\theta_{12}=33.68^\circ$ and $\delta_{\textrm{CP}}=180^\circ$ (the top-left framed subplot). Changing the above order of rotation matrices $R_x,R_y,R_z$, we can also find angles leading to the same range of experimental values as for $U_{\texttt{PMNS}}$, however, corresponding angles are completely different, for example 
\begin{eqnarray}U&=&R_x(43.3^\circ) R_y(8.56^\circ)R_z(33.68^\circ)=R_x(37.64^\circ)R_z(10.25^\circ)R_y(33.26^\circ)
\nonumber 
\\&=&R_y(42.70^\circ)R_x(11.69^\circ)R_z(25.70^\circ)=R_y(18.58^\circ)R_z(45.68^\circ)R_x(29.76^\circ) \label{eul} \\
&=&R_z(33.17^\circ)R_x(39.63^\circ)R_y(30.71^\circ)=R_z(35.36^\circ)R_y(28.06^\circ)R_x(21.17^\circ) \nonumber.
\end{eqnarray}
 
In other words,  the experimentally determined interval mixing matrix, say $\vosc$, can be covered by the range of mixing angles in six different ways. Using the 1$\sigma$ experimental allowed range~\cite{esteban2025nufit}, one can estimate $U \to  \vosc $ as an interval matrix, given by \begin{eqnarray}
  \label{ranges}
 U \to  \vosc = 
&& \begin{pmatrix}
    0.817 \div 0.831 & 
    0.535 \div 0.558 &
    -0.150 \div -0.146
    \\
    -0.313 \div -0.273 &
    0.615 \div 0.657 &
    0.686 \div 0.738
    \\
    0.471 \div 0.498 &
    -0.569 \div -0.519 &
    0.658 \div 0.713
    \end{pmatrix}\;,  
\end{eqnarray}
where we have considered $\delta_\textrm{CP} = 180^\circ$. 
Similarly, for $\delta_{\rm CP}=0^{\circ}$ CP-conserving scenario,   the sign of the `13' element in $\vosc$ changes, keeping signs of all other elements intact. 

Therefore, for each order of rotations in Eq. (\ref{Euler}), the obtained range of mixing angles that can cover $\vosc$ in Eq. (\ref{ranges}) will be different. It is interesting to note that the commonly used choice $R_xR_yR_z$ leads to the mixing angles with best fit values, which are almost maximal for $\theta_{23}$ and large $\theta_{12}$. Experimentally, we just measure Eq. (\ref{ranges}) and after decades of observation we find that $|(\vosc)_{e1}|^2 > |(\vosc)_{\mu 1}|^2>|(\vosc)_{\tau 1}|^2$ which follows from the specific order of rotation $R_xR_yR_z$.  It leads to many speculations on possible flavor symmetries behind \cite{Chauhan:2023faf}. 

In \cite{Denton:2020igp}, permutations of Euler matrices were considered, showing numerically that they are qualitatively equivalent. This can be understood as {\it mixing angles are parameters of the $U$ matrix, not physical observables, fitted experimentally}.  So there is nothing special in choosing $U_{\rm PMNS}$ as the neutrino mixing parametrization.  However, a specific representation matters as many theoretical considerations and mass/mixing models are built based on this choice. 

In what follows, we propose an order-independent parameterization. This can be done by considering the SO(3) group, where we can perform one direct rotation instead of 3 in arbitrarily chosen orthogonal directions.

\subsection{The proposed SO(3) order independent parameterization}
The continuous SO(3) family symmetry is one of the most widely used Lie groups in mathematics and physics~\cite{Hall:2015xtd, Georgi:2000vve}. In order to impose SO(3) symmetry directly on the neutrino mixing matrix, let us briefly discuss the  SO(3) group representation. SO(3) can be generated by three generators, namely, $G_x, G_y, G_z$. In the fundamental three-dimensional space, 
the standard antisymmetric SO(3) generators can be written as~\cite{Georgi:2000vve}
\begin{equation}\label{eq:generators}
G_x = \left(
\begin{array}{ccc}
 0 & 0 & 0 \\
 0 & 0 & -1 \\
 0 & 1 & 0 \\
\end{array}
\right),\qquad
G_y =\left(
\begin{array}{ccc}
 0 & 0 & 1 \\
 0 & 0 & 0 \\
 -1 & 0 & 0 \\
\end{array}
\right),
\qquad 
G_z = \left(
\begin{array}{ccc}
 0 & -1 & 0 \\
 1 & 0 & 0 \\
 0 & 0 & 0 \\
\end{array}
\right) , 
\end{equation}
{where $G^{T}_i=-G_i$ and the commutator of the generators can be written as 
\begin{align}
    \left[G_i, G_j\right]=\epsilon_{ijk}G_k, \quad i,j,k=x,y,z. 
    \label{commut}
\end{align}
This is a unique construction driven by Levi-Civita antisymmetric epsilon. {It is also invariant under basis changes, ensuring that the results are consistent regardless of the basis.} From Eq. (\ref{eq:generators}) we find that, 
\begin{eqnarray}
    G_z^2=-\left(\begin{array}{ccc}
 1 & 0 & 0 \\
 0 & 1 & 0 \\
 0 & 0 & 0 \\
\end{array}
\right) \equiv -P; G^3_z=G_z G^2_z=-G_z; G^4_z=G_z G^3_z=-G^2_z=P, G^6_z=-P.
\end{eqnarray}
The three-dimensional rotations do not commute, consequently, the SO(3) rotation group in three dimensions is not Abelian. It is straightforward to show, connecting  the ordered Euler rotation matrices  $R_x(\theta_{23}), R_y(\theta_{13},\delta_{\rm CP}=180^{\circ}), R_z(\theta_{12})$ with exponents of SO(3) generators $G_x,G_y,G_z$,  that   
\begin{eqnarray}
    U&=&  
    e^{ -\theta_{23} G_x}e^{-\theta_{13}G_y}e^{-\theta_{12}G_z}, \\
    &=& \begin{pmatrix}
           1 & 0 & 0\\
           0 & c_{23} & s_{23}\\
           0 & -s_{23} & c_{23}           
        \end{pmatrix}
         \begin{pmatrix}
           c_{13} & 0 & -s_{13}\\
           0 & 1 & 0\\
           s_{13} & 0 & c_{13}           
        \end{pmatrix}
        \begin{pmatrix}
           c_{12} & s_{12} & 0\\
           -s_{12} & c_{12} & 0\\
           0 & 0 & 1           
        \end{pmatrix},\\
    &=& \begin{pmatrix}
c_{12} c_{13} & s_{12} c_{13} & -s_{13}  \\
-s_{12} c_{23}+ c_{12} s_{23} s_{13}  & 
c_{12} c_{23} + s_{12} s_{23} s_{13}  & 
s_{23} c_{13} \\
s_{12} s_{23} +c_{12} c_{23} s_{13}  & 
-c_{12} s_{23} + s_{12} c_{23} s_{13}  & 
c_{23} c_{13}
\end{pmatrix},
\end{eqnarray}
corresponds to the $U_{\rm PMNS}$ in Eq. (\ref{upmns1}) when fixing the Dirac CP phase to be $\delta_{\rm{CP}}=180^\circ$\footnote{Similarly, for $\delta_{\rm CP}=0^\circ$ the PMNS equivalent mixing matrix can be obtained from the corresponding expression $U=e^{ -\theta_{23} G_x}e^{+\theta_{13}G_y}e^{-\theta_{12}G_z}$.}. This is so as the nonzero elements of the SO(3) generator $G_y$ have the opposite sign to the elements of $G_x$ and $G_z$, see Eq. (\ref{eq:generators}). It can also be seen from the opposite sign of the sine of the angle in the rotation matrix $e^{-\theta_{13} G_y}$ when comparing with $e^{-\theta_{23} G_x}$ and $e^{-\theta_{12} G_z}$  (clockwise rotation in all $x,y,z$ directions driven by antisymmeric Levi-Civita tensor with cyclic order $23,31,12$ of generators in Eq. (\ref{commut})).  Other combinations, such as  $e^{ -\theta_{23} G_x}e^{-\theta_{12}G_z}e^{-\theta_{13}G_y}$ or $e^{-\theta_{12}G_z} e^{ -\theta_{23} G_x}e^{-\theta_{13}G_y}$ will imply different values of the neutrino mixing matrix angles, not reflecting maximal mixings, see Eq.~(\ref{eul}). 

In order to get rid of such order dependence for multiplication of group matrices, we introduce an \emph{order independent parameterization} of rotations in a plane perpendicular to $\vec{\Theta} (\theta_x, \theta_y, \theta_z)$  through the matrix exponent.  Using the three SO(3) generators in Eq.~(\ref{eq:generators}) and the three angles defined here, one can write their inner product as  
\begin{eqnarray}\label{param00}
   -\vec{\Theta}.\vec{G}=-\theta_{x}G_x-\theta_{y}G_y -\theta_{z} G_z=\left(
\begin{array}{ccc}
 0 & \theta_z & -\theta_y \\
 -\theta_z & 0 & \theta_x \\
 \theta_y & -\theta_x & 0 \\
\end{array}
\right). 
\end{eqnarray}
This matrix is traceless, anti-symmetric.  Hence, the SO(3) rotation operator can be written as 
\begin{eqnarray}
U_{\rm SO3}&=&\exp\left(
\begin{array}{ccc}
 0 & \theta_z & -\theta_y \\
 -\theta_z & 0 & \theta_x \\
 \theta_y & -\theta_x & 0 \\
\end{array}
\right)\\
&=&
{\left(
\begin{array}{ccc}
 \frac{\theta _x^2+\cos \theta  \left(\theta _{y }^2+\theta _{z }^2\right)}{\theta ^2} & \frac{\theta  \theta _{z } \sin \theta -\theta _x \theta _{y} (\cos \theta -1)}{\theta ^2} & -\frac{\theta _x \theta _{z } (\cos \theta -1)+\theta  \theta _{y } \sin \theta }{\theta ^2} \\
 -\frac{\theta _x \theta _{y } (\cos \theta -1)+\theta  \theta _{z } \sin \theta }{\theta ^2} & \frac{\cos \theta  \left(\theta _x^2+\theta _{z }^2\right)+\theta _{y }^2}{\theta ^2} & \frac{\theta  \theta _x \sin \theta -\theta _{y } \theta _{z } (\cos \theta -1)}{\theta ^2} \\
 \frac{\theta  \theta _{y } \sin \theta -\theta _x \theta _{z } (\cos \theta -1)}{\theta ^2} & -\frac{\theta  \theta _x \sin \theta +\theta _{y } \theta _{z} (\cos \theta -1)}{\theta ^2} & \frac{\cos \theta  \left(\theta _x^2+\theta _{y }^2\right)+\theta _{z }^2}{\theta ^2} 
\end{array}
\right)}.\label{param2}
\end{eqnarray}
where $\theta=\sqrt{\theta_{x}^2+\theta_{y}^2+\theta_{z}^2}$\footnote{Supporting Mathematica \cite{Mathematica} file with derivation of the $U_{\rm SO3}$ structure in Eq.~(\ref{param2}) can be found in \cite{SO3www}.}.   The salient feature of the $U_{\rm SO3}$ mixing matrix is that it is orthogonal and unimodular, and hence represents a rotation.  Physical consequence of the condition $\delta_\text{\rm CP}=180^\circ$  in $U_{\rm SO3}$ is that the matrix is real, so we conjecture that the neutrino oscillations preserve CP symmetry \cite{Bilenky:1987ty, delAguila:1996ex, Gluza:1995ky}.  To determine the angles $\theta_x,\theta_y,\theta_z$ for $U_{\rm SO3}$ in Eq.~(\ref{param2}),  we used  NuFIT-6.0 data~\cite{esteban2025nufit}  and get at 1$\sigma$
\begin{eqnarray}
   &&   \theta_{x}=43.8^{\circ}\pm 2.1 ^{\circ}, \ \quad \theta_{y}=21.7^{\circ}\pm 0.6^{\circ},  \quad \theta_{z}=28.3^{\circ}\pm 0.8^{\circ}
   \label{angles2}
\end{eqnarray}  
with $\theta \approx 56.5^\circ$. Similarly, for $\delta_{\rm CP}=0^{\circ}$ the corresponding allowed ranges for $\theta_x,\theta_y,\theta_z$ are found to be 
\begin{eqnarray}
   &&  
   \theta_{x}=46.5^{\circ}\pm 2.4^{\circ},  \quad  \theta_{y}=5.4^{\circ}\pm 0.8 ^{\circ}, \quad  \theta_{z}=35.1^{\circ}\pm 0.7 ^{\circ}
     \label{angles2_0}
\end{eqnarray}  
with $\theta \approx 58.5^\circ$. See Fig.~\ref{figso3} for visualization of the order-independent single SO(3) rotation against the situation in the $U$  case in Fig.~\ref{Euler}.
 \begin{figure*}[h]
    \centering
        \includegraphics[width=.35\textwidth]{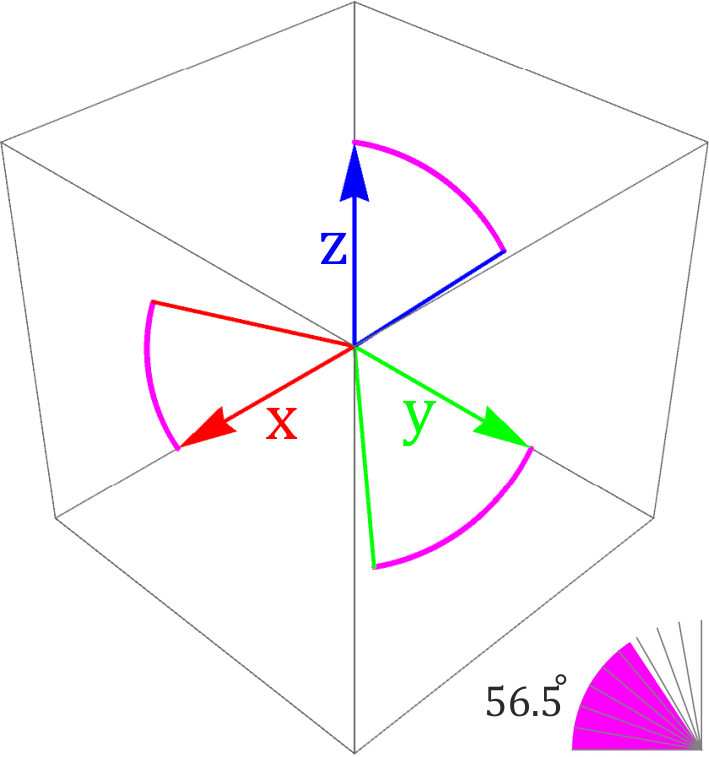}
        \caption{{The proposed direct rotation by $\theta=\sqrt{\theta_{x}^2+\theta_{y}^2+\theta_{z}^2}\simeq 56.5^{\circ}$ for $\delta_{\rm CP}=180^{\circ}$.  
         The arc plot 
        visualises values of the projected rotations for the determined neutrino mixing angles in Eq. (\ref{angles2}) 
        }.}
       \label{figso3}
\end{figure*}
\begin{figure*}[h!]
  \centering 
        \includegraphics[width=16cm]{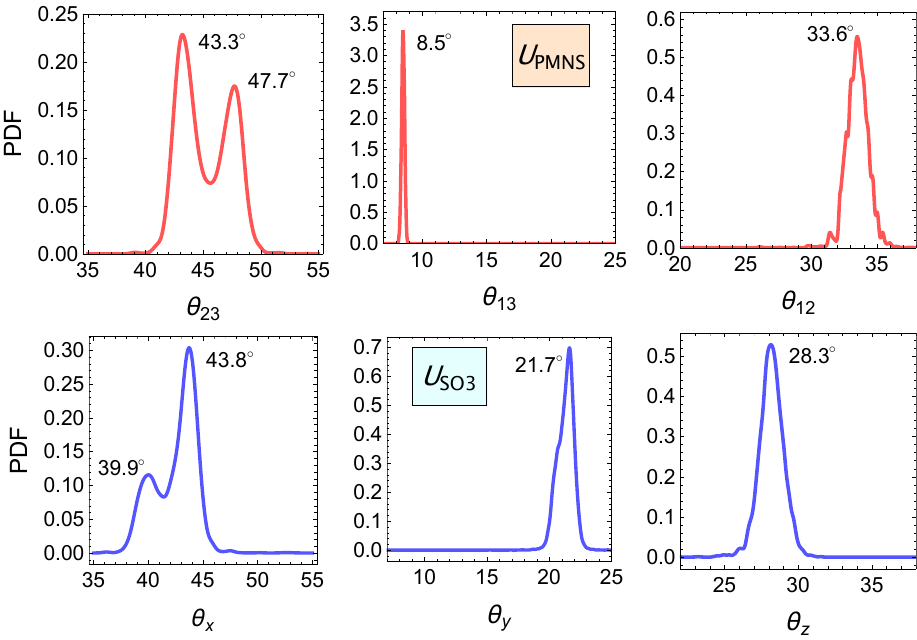}
        \caption{Probability densities (PDF) based on  NuFIT-6.0 data~\cite{esteban2025nufit} for the original PMNS parameterization with any values of $\delta_{\rm CP}$ ($\theta_{23,13,12}$, top plots) and the proposed order-independent SO(3) parameterization, so with $\delta_{\rm CP}=180^\circ$ ($\theta_{x,y,z}$, bottom plots). 
     Values inside each plot are angles for PDF maxima.   
        }
       \label{thetas}
\end{figure*}
Fig.~\ref{thetas} compares probability densities for the mixing angles $\theta_{23}$, $\theta_{13}$, $\theta_{12}$  and $\theta_{x}$, $\theta_{y}$, $\theta_{z}$  
derived using the NuFIT-6.0 data for PMNS and SO(3) parameterizations respectively.  Clearly, the $U_{\rm SO3}$ matrix is expressed with values of the mixing angles that are closer to each other compared to the PMNS mixing for $\delta_{\rm CP}=180^\circ$. Whereas, for $\delta_{\rm CP}=0^\circ$ the mixing angles acquire near maximal values which lie in close proximity to PMNS mixing. Therefore, the SO(3) parameterization allows 
two distinct lepton mixing patterns with different phenomenological implications. 

Specifically, the $U_{\rm SO3}$ parameterization, for $\delta_{\rm CP}=180^\circ$,  gives smaller $\theta_x$ and $\theta_z$ angles than  $\theta_{23}$ and $\theta_{12}$ angles in the $U_{\rm PMNS}$ case. $\theta_{y}$ is much larger than $\theta_{13}$. What can be interesting for model builders, $\theta_x$ lies entirely in the first quadrant.  Unsurprisingly, as $U_{\rm SO3}$  matches with the experimentally observed interval matrix $\vosc$, following Eq.~(\ref{class}), the flavor contributions in each mass eigenstate in both  $U_{\rm SO3}$ and $U_{\rm PMNS}$ remain identical. The exponential form in Eq.~(\ref{param2}), built from Eq.~(\ref{param00}), is in general basis-dependent. Therefore, under a basis transformation, the magnitude of $\theta_{x}, \theta_{y}$ and $\theta_{y}$ will change.  The rotation itself, however, will remain the same, and all such forms are related by similarity transformation. Since similarity transformations preserve intrinsic quantities such as the rotation angle, the physical content of the transformation remains unchanged.

\section{Discussion of  the $U_{\rm SO3}$  neutrino mixing setup}

As we can see, the $U_{\rm SO3}$ parameterization leads to completely different mixing angles obtained in Eq. (\ref{angles2}) compared to the $U_{\rm PMNS}$ case in Eq.  (\ref{eq:oscdata}). In particular, for $\delta_{\rm CP}=180^\circ$,  there is no small angle which would correspond to $\theta_{13}$ in the PMNS case.  
In the PMNS case, the $\theta_{13}$ angle is connected with the (1,3) element of the the mixing matrix $U$ in Eq. (\ref{class}), which is the smallest experimentally determined oscillation mixing matrix element, of the order of 0.15, see $\vosc$ in Eq. (\ref{ranges}). In the PMNS parameterization the (1,3) element reads $s_{13}e^{-i\delta_{\rm CP}}$, thus $\theta_{13}$ {\it is considered to be} small. On the other hand, elements in the proposed matrix $U_{\rm SO3}$ are  a combination of all angles $\theta_{x},\theta_{y},\theta_{z}$, leading to relatively large values of all three mixing angles for $\delta_{\rm{CP}}=180^\circ$.  Neutrino oscillation probabilities depend on the combination of squared $U$ elements in Eq. (\ref{class}) from which ranges of $\vosc$ values can be derived. Evidently,  single rotation in the SO(3) space, involving three large mixing angles $\theta_{x},\theta_{y},\theta_{z}$ recover ranges of $\vosc$ correctly.

A natural question is, what can we gain by introducing a new  $U_{\rm SO3}$  parameterization? Well, apart from the obvious advantage of having an order-independent rotational mixing framework, the $U_{\rm SO3}$ parameterization with strict CP-conserving cases lead to intriguing implications. Firstly,  a fixed value of $\delta_{\rm CP}$ narrows predictions for the  tritium beta decay~\cite{Formaggio:2021nfz} $m^2_{\beta}$ parameter, defined as 
\begin{align}
    m^2_{\beta}
    = \frac{\sum_i m^2_i |U_{ei}|^2}{\sum_i |U_{ei}|^2}
    =\sum_i m^2_i |U_{ei}|^2 \, , \label{beta}
\end{align}
{\it assuming $U$ is orthogonal}\footnote{In general, non-unitary $U$ mixing with different numbers of additional neutrino states can be disentangled and classified the best way by singular values of $U$ \cite{Bielas:2017lok,Flieger:2019eor}, see also recent \cite{Blennow:2025qgd} for potential misconceptions in description of non-unitary neutrino oscillations.}.  Secondly, the effective mass parameter $ m_{\beta\beta}$, which appears in the neutrinoless double beta decay, is defined in the following way\footnote{Other rare lepton flavour violating processes like $\mu \to e\gamma$ are not sensitive to the light neutrino spectrum \cite{Cheng:1984vwu}.}~\cite{Dolinski:2019nrj}
\begin{equation}
  m_{\beta\beta} = \Big| \sum_i m_i U_{ei}^2 \Big|  , \label{2beta}
\end{equation}
also gets significantly constrained with vanishing Majorana phases, as justified in section \ref{standard}. This can be seen in Fig.~\ref{fig:mbb} which presents results for NO hierarchy (IO is excluded with $\delta_{\rm CP}=180^\circ$). The elements of $U$ in Eq. (\ref{beta}) and Eq. (\ref{2beta}) where replaced by mixing matrix elements within the PMNS and  SO(3) frameworks, in Eq. (\ref{upmns1}) and Eq. (\ref{param2}), respectively.  At present, the best direct limit on $m_\beta$ comes from the tritium beta decay experiment KATRIN: $m_{\beta} < 0.8$~eV at 90\% CL~\cite{KATRIN:2021uub}, with projected sensitivity down to $m_{\beta} < 0.2$~eV at 90\% CL~\cite{KATRIN:2021dfa}. The future Project~8 experiment using the Cyclotron Radiation Emission Spectroscopy (CRES) technique is expected to reach a sensitivity for $m_{\beta}$ down to 0.04 eV~\cite{Project8:2022wqh}. As we can see in the top panel of Fig.~\ref{fig:mbb},  predicted values of $m_{\beta}$ (against lightest neutrino mass $m_1$) get significantly constrained by the SO(3) mixing hypotheses.  Here, the green (dark-red) shaded region represents 1$\sigma$ allowed region considering  SO(3) mixing hypotheses with  $\delta_{\rm{CP}}= 0^\circ(180^\circ)$. Interestingly, following Eq. (\ref{beta}), here we find distinguishable prediction for $m_{\beta}$ for two different CP-conserving cases following their distinct ranges for $\theta_y$ and $\theta_z$ given in Eq. (\ref{angles2}) and Eq. (\ref{angles2_0}), respectively. 

Similarly, the present best upper limit on $m_{\beta\beta}$ comes from the KamLAND-Zen experiment using $^{136}{\rm Xe}$: $m_{\beta\beta}<0.036 - 0.156$ eV at 90\% CL~\cite{KamLAND-Zen:2022tow}, where the range is due to the nuclear matrix element (NME) uncertainties. Several next-generation experiments are planned with different isotopes~\cite{Adams:2022jwx}, with ultimate discovery sensitivities to $m_{\beta\beta}$ down to 0.005 eV \cite{KamLAND-Zen:2022tow,GERDA:2020xhi,nEXO:2021ujk}. 
\begin{figure}[h!]
    \centering
\includegraphics[width=0.78\linewidth]{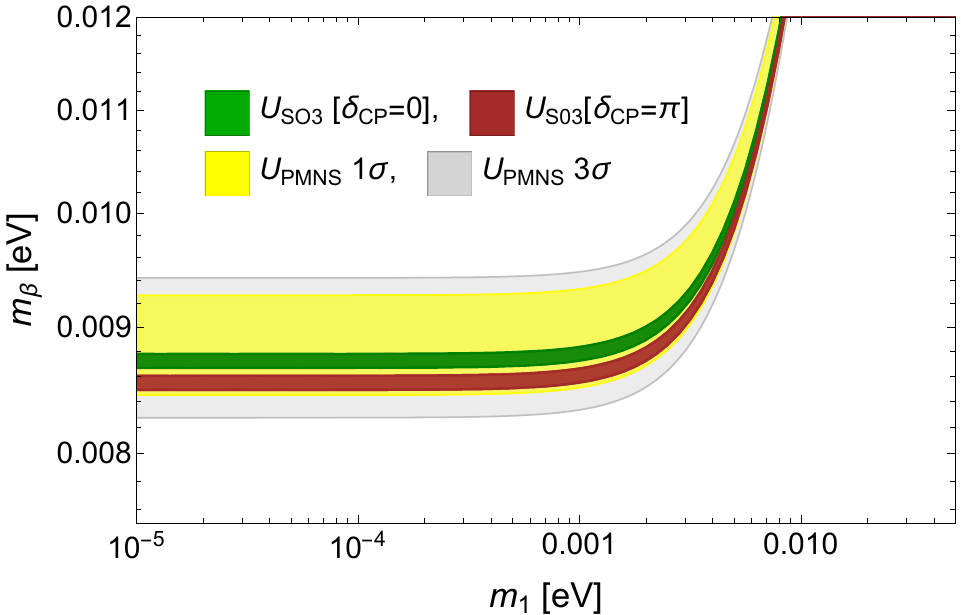}
\\
\includegraphics[width=0.78\linewidth]{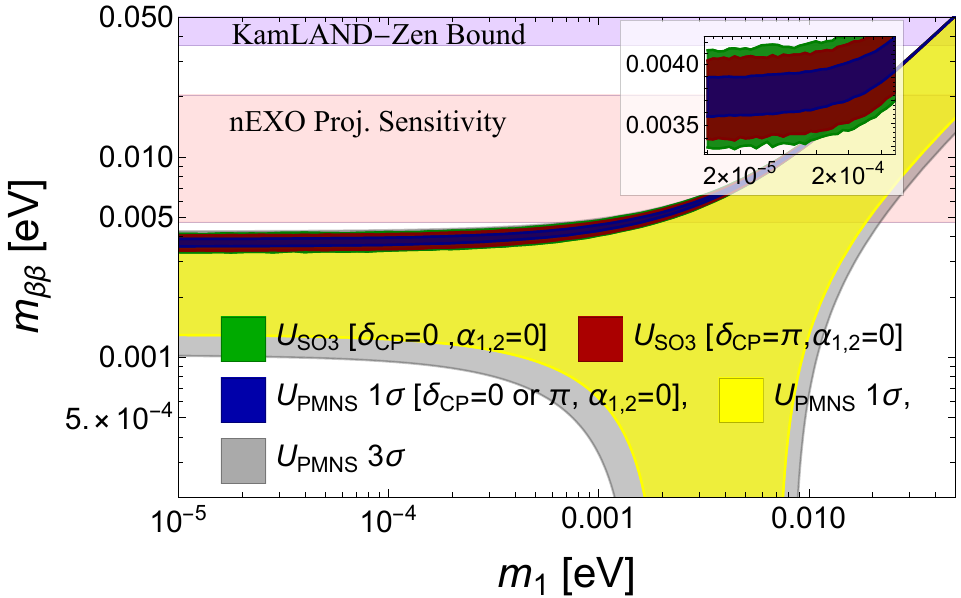}
    \caption{
    Prediction for $m_{\beta}$ (top) 
    and $m_{\beta\beta}$ (bottom) against the lightest neutrino mass ($m_1$) with 
    $\delta_{\rm CP}=0^\circ$ and $180^\circ$ for NO. In the both  panels, the green (dark-red)  shaded region represents the SO(3) prediction for $\delta_{\rm CP}=0^\circ$  ($180^\circ$). The yellow and gray shaded regions represent the 1$\sigma$ and 3$\sigma$ allowed region~\cite{esteban2025nufit} using $U_{\rm PMNS}$ parameterization. The dark blue patch in the bottom panel shows the 1$\sigma$ $U_{\rm PMNS}$ allowed regions with $\delta_{\rm CP}=0^{\circ}(180^{\circ})$ and $\alpha_{1,2}=0$.}
    \label{fig:mbb}
\end{figure}

In the bottom panel of Fig. (\ref{fig:mbb}), the $m_{\beta\beta}$ prediction is given against the lightest neutrino mass $m_1$ for the $U_{\rm SO3}$  mixing framework, shown by the green (dark-red) shaded region for $\delta_{\rm CP}=0^\circ({180^\circ})$. Here, due to the CP-conserving values of the associated phases, the SO(3) prediction lies at the maximum extreme edge of the standard prediction, and very small $m_{\beta\beta}$ values are not possible due to vanishing Majorana CP phases to include destructive contributions, which explains the cancellation dip. This is also evident by the dark blue shaded region in the bottom panel, which shows the 1$\sigma$ $U_{\rm PMNS}$ allowed region with $\delta_{\rm CP}=0^{\circ}(180^{\circ})$ and $\alpha_{1,2}=0$, yielding similar prediction for the SO(3) construction. Therefore, unsurprisingly, the SO(3) mixing hypothesis puts a stringent constraint on $m_{\beta\beta}$ due to strong restrictions on the associated CP phases. Unlike the distinctive $m_\beta$ prediction in the top panel, the predictions for $m_{\beta\beta}$ for CP-conserving cases  $\delta_{\rm CP}=0^{\circ}~{\rm and}~(180^{\circ})$ falls in the similar ballpark and their slight difference has been highlighted in the inset of the bottom panel.
In both panels, the gray and yellow shaded regions represent the 1$\sigma$ and 3$\sigma$ allowed ranges, respectively, considering $U_{\rm PMNS}$ parameterization.

\section{Conclusions and outlook}\label{sec:conc}

Solar and atmospheric neutrino oscillations restricted to the mass-flavor mixing matrix $U$ between two pairs of neutrino species lead to the prediction of two (almost) maximal mixings. Thus, the choice of the unitary three-dimensional $U_{\rm PMNS}$ parameterization emerges as a product of $R_x R_y R_z$ neutrino flavor rotations in a natural way with (almost) tribimaximal mixing. With large $\theta_{12}$, maximal mixing for $\theta_{23}\sim \pi/4$ and relatively small $\theta_{13}$ angle this situation triggers theoretical speculations on possible symmetries behind, with phenomenological consequences \cite{Chauhan:2023faf}. However, in general, the starting point for neutrino mixing and derived probabilities is the complete three-dimensional mass-flavor mixing matrix in Eq.~(\ref{class}), which can be derived from global experimental fits. Thus, we first express experimentally established neutrino mixings in the form of an interval matrix $\vosc$.  Then we note that the $U_{\rm PMNS}$ parameterization is one of six possible multiplication of Euler matrices $R_x, R_y, R_z$, each leading to different values of $\theta_{12}$, $\theta_{23}$  and $\theta_{13}$ angles {\it for the same experimentally established elements of the matrix $U$ in} Eq. (\ref{class}). From a theoretical point of view, all six $R_x, R_y, R_z$ permutations and corresponding parameterizations are equivalent (the decomposition of a unitary matrix into selected basis matrices for the unitary group is not unique). 

To be independent of this order dependence, we propose a parameterization of the 3-dimensional $U$ matrix based on the SO(3) group, which corresponds to one general rotation in 3-dimensional space. Naturally, the representation of the SO(3) generators is real. To harmonize the signs of elements of the SO(3) generators and related commutator relation with the neutrino mixing matrix that follow, we infer $\delta_{\rm{CP}}=180^\circ$.  
In this way, we get the $U_{\rm SO3}$ parameterization for mixing of three known neutrinos, for which determined mixing angles $\theta_{x}, \theta_{y}, \theta_{z}$ differ from the $U_{\rm PMNS}$ parameterization case.  The consequence of the SO(3) parameterization is that we predict CP-conservation in neutrino oscillations with $\delta_{\rm{CP}}=180^\circ$ and as all three determined mixing angles $\theta_{x}, \theta_{y}, \theta_{z}$ are substantial, we also get tighter constraints on $m_{\beta}$ and  $m_{\beta\beta}$ for the SO(3) case compared to the standard $U_{\rm PMNS}$ predictions.  

As an outlook, we think that: 
\begin{enumerate} 
\item The proposed  $U_{\rm{SO3}}$ order-independent neutrino mixing parameterization is well suited for independent  CP-conserving neutrino oscillation experimental analysis. 
It will be interesting to explore the structure of the neutrino oscillation probabilities arising from the constructed $U_{\rm{SO3}}$ mixing matrix and their implications for neutrino data analysis.
\item 
Both  T2K and NO$\nu$A disfavours inverted ordering with $\delta_{\rm{CP}}=180^\circ$. In the framework of the SO(3) parameterization, we remain with the normal mass ordering for both of the CP-conserving scenarios and in inverted mass ordering for $\delta_{\rm{CP}}=0^\circ$.

\item The SO(3) mixing hypothesis presented here can accommodate both of the CP-conserving values. It yields near maximal mixing for $\delta_{\rm CP}=0^{\circ}$, whereas relatively large, `democratic' values of the mixing angles  $\theta_{x},\theta_{y},\theta_{z}$. This demands extension of the SM gauge symmetry and may point toward specific flavor symmetry-breaking patterns that naturally generate the SO(3) mixing hypothesis.
\item The distinct predictions of the SO(3) framework with tighter constraints on $m_{\beta}$ and $m_{\beta\beta}$ provide concrete targets for upcoming experiments. In particular, the predicted values for $m_{\beta\beta}$ within SO(3) lie at the edge of current PMNS predictions, making them testable with next-generation neutrinoless double beta decay searches (nEXO, LEGEND). 

\item The fixed $\delta_{\rm{CP}}=180^\circ$ prediction eliminates CP violation in the lepton sector, which could have implications for a wide variety of leptogenesis scenarios. For example, with hierarchical right-handed neutrinos, the lepton asymmetry vanishes for vanilla leptogenesis. Flavor effects and other non-standard effects can lead to successful leptogenesis~\cite{Davidson:2008bu, Drewes:2016jae, Dev:2017wwc}.   
\end{enumerate}
\vspace*{-.cm}
The connection between spatial rotations and flavor mixing may suggest a deeper understanding of the leptonic flavor puzzle. Although we do not advocate SO(3) as a fundamental flavor symmetry, the fact that current neutrino mixing data can be accurately described by a single SO(3) rotation is intriguing. This geometric structure might serve as an interesting boundary condition for future model building.  Finally, as conjectured by one of the authors of the present work in \cite{duda2021}, neutrinos can be viewed as some spatial field rotations (topological solitons), and it is also worth exploring at a deeper level the connection between SO(3) neutrino flavour mixings and the description of neutrinos as topological objects (work in progress).

\section*{Acknowledgments}

We thank Monojit Ghosh, Wojciech Flieger and Federico Sánchez for useful remarks. This work has been supported in part by the Polish National Science Center (NCN) under grant 2020/37/B/ST2/02371, the Research Excellence Initiative of the University of Silesia in Katowice and the Swiss National Science Foundation (SNSF) under grant MAPS IZ11Z0\_230193.

\bibliography{bibliography,biblio2,ref,ref1}
	\bibliographystyle{utphys}
\end{document}